\documentclass[aps,prb,twocolumn,superscriptaddress]{revtex4-2}
\usepackage{amsmath}
\usepackage{graphicx}
\usepackage{subcaption}
\usepackage{xcolor}
\usepackage{ragged2e}
\usepackage{amssymb}
\usepackage{mathtools}

\DeclareMathOperator{\tr}{tr}

\DeclareMathOperator{\arccosh}{arccosh}

\begin{document}

	%\title{SU(3) symmetry breaking induces superconducting pairing\\ in the repulsive Hubbard model}
	\title{ Dominant Excitonic  Superconductivity in a Three-component Hubbard Chain}

	\author{Sheng Chen}
	\affiliation{School of Physics, Sun Yat-Sen University, Guangzhou, 510275, China}
	\affiliation{School of Science, Great Bay University, Dongguan, 523000, China}

	\author{Qiao Yang}
	\affiliation{School of Science, Great Bay University, Dongguan, 523000, China}
	\affiliation{School of Physics, University of Science and Technology of China,Anhui 230026, China}
	
	\author{Wéi Wú}
	\email{wuwei69@mail.sysu.edu.cn}
	\thanks{Corresponding author} 
	\affiliation{School of Physics, Sun Yat-Sen University, Guangzhou, 510275, China}

	\author{Fadi Sun}
	\email{fadisun@gbu.edu.cn} 
	\thanks{Corresponding author} 
	\affiliation{School of Science, Great Bay University, Dongguan, 523000, China}

	\begin{abstract}
		Understanding superconductivity emerging from repulsive fermions remains a major challenge in condensed matter physics. In this paper, we investigate the pairing tendencies in a one-dimensional, three component repulsive Hubbard model, using the density matrix renormalization group  method. At half-filling, the system exhibits density wave ground state due to strong Hubbard repulsions. Upon doping, we find that Cooper pairs can emerge, whose fluctuations predominate the long-range physics   in the system across a wide parameter range. The effective attractions between Cooper pairs are mediated by the particle-hole fluctuations in the third non-pairing component, resembling an excitonic mechanism of superconductivity. The coexistence of multiple density waves and superconductivity at different fermion  fillings is explored. We also present an analytical study of the pairing mechanism in both weak and strong coupling limits. Our results provide a new perspective for understanding and exploring unconventional superconductivities in strongly correlated  fermionic systems.
		
	\end{abstract}
	
	\maketitle

	\section{Introduction}
	
	The Hubbard model is widely regarded as the prototypical  model for investigating many strongly correlated electronic systems~\cite{arovas2022hubbard,tasaki1998hubbard}. It plays a particularly pivotal role in the study of high-temperature superconductivity~\cite{keimer2015quantum,maier2000d,qin2020absence,maier2005systematic}, where two-dimensional (2D) Hubbard model successfully describes diverse phenomena such as d-wave pairing~\cite{maier2005systematic,liu2025interplay}, pseudogap~\cite{wu2018pseudogap,wu2020effect}, and strange metal states~\cite{wu2022non} in cuprates. The Hubbard model provides a fundamental basis for understanding electron correlation effects by encoding the competition between kinetic energy $\epsilon(k)$ and Coulomb repulsion $U$  of electrons. 
	While the attractive Hubbard model ($U < 0$) is often tractable with quantum Monte Carlo simulations where superconducting states can be readily established ~\cite{cheng2024drastic,toschi2005pairing}, the repulsive cases ($U > 0$) present a far greater challenge for analytical or numerical studies.  For $U >0$, it is  believed that electron pairing in Hubbard models may emerge through same intricate mechanisms, such as the short-ranged  anti-ferromagnetic correlations~\cite{clay1999absence,tam2014dominant,dong2022mechanism,roig2022revisiting,raghu2010superconductivity}. However, whether true long-range superconductivity exists in 2D repulsive Hubbard model is yet a key open question under debate~\cite{qin2020absence}.

In one-dimensional (1D) or quasi-one-dimensional systems, the superconducting properties of Hubbard model are much better understood.
	It has been shown that  a purely repulsive on-site   Hubbard $U >0$  does not lead to dominant pairing %{[cite H.Q.Lin]
	~\cite{lin1997phase,clay1999absence}. Instead, a significant nearest neighbour repulsion $V$ is typically required to drive long-ranged superconducting fluctuations~\cite{lin1997phase,lin2000broken}.
	In quasi-one-dimensional $n-$leg Ladders systems, density matrix renormalization group (DMRG) study demonstrate that  when the competing charge-density-wave (CDW) order is suppressed by  next-nearest-neighbor hopping~\cite{jiang2024ground,jiang2021ground,jiang2020ground,jiang2019superconductivity,lu2024emergent,gong2021robust} ,   a Luther-Emery liquid phase can be stabilized, where superconductivity and CDW  coexists.

	In this work, we study the dominant superconducting fluctuations in  a modify $\mathrm{SU(3)}$ repulsive Hubbard model on one dimensional chain, using analytical analysis and DMRG calculations.
	%This model was proposed in Ref., where it is argued that electron pairing due to strong-coupling Kohn-Luttinger mechanism can emerge.
	Previous research on the interacting SU($N$) fermions has mainly focused on the metal-insulator transition~\cite{inaba2010mott,feng2023metal,hermele2009mott,yanatori2016finite,sotnikov2014magnetic,hafez2019competing} in the  SU($N$) Hubbard model, or spin dynamics in the $t\text{-}J$~\cite{bohler2025magnetic,schlomer2024subdimensional} and Heisenberg models~\cite{assaad2005phase}. 
	Here by breaking the SU($3$) symmetry of the Hubbard interactions in a three component Hubbard model, we reveal strong effective attraction between two of the fermionic components with reduce Hubbard repulsion $U^{\prime}$. This  attraction is driven by the particle-hole fluctuations in the third component, which interacts repulsively with the other  two components via  a strong Hubbard $U > 2U^{\prime}$. Our model realize a long-proposed form of unconventional supercoducvtivity
	mediated by excitonic effects~\cite{little1964possibility,davis1976proposed,crepel2022spin,singh2022unconventional}, which is fundamentally  different from the conventional magnetically driven  superconductivity 
	in the doped SU(2) Hubbard model.

	Our paper is organized as follows. In Sec.II we introduce the modified SU(3) Hubbard model and the method used in this article. In Sec. III we study the density wave states and present analytical analysis  at half-filling, then we present superconducting phase diagram and tackle the coexistence of DW and SC. Finally, in Sec. IV we discuss the potential realization of our model and provide a conclusion.

	\section{MODEL AND METHOD}
	
	%We use the DMRG~\cite{white1992density,schollwock2005density} method to investigate the ground-state properties of 
	We consider a one-dimensional, three-component fermionic Hubbard model, defined by the Hamiltonian,
	\begin{align}\label{eq1}
		H ={} & -t \sum_{\langle i,j \rangle, \alpha} 
		\left( \hat{c}^\dagger_{i,\alpha} \hat{c}_{j,\alpha} + \textit{h.c.} \right)\notag \\
		&+ \sum_{i,\alpha \neq \beta}  U_{\alpha, \beta}  \hat{n}_{i,\alpha} \hat{n}_{i,\beta}  \notag \\
		&- \mu \sum_{i, \alpha} \hat{n}_{i,\alpha}
		-  \sum_i \epsilon_{\alpha}  \hat{n}_{i, \alpha} 
	\end{align}
	
	where $\hat{c}^\dagger_{i,\sigma}(\hat{c}_{i,\sigma} )$ is the creation (annihilation) operator for a fermion with spin flavor $\alpha$ ( $\alpha$=1,2,3 ) at site $i$, and $\hat{n}_{i,\alpha} = \hat{c}^\dagger_{i,\alpha} \hat{c}_{i,\alpha}$ is the corresponding number operator. The hopping amplitude between nearest-neighbor sites $i$ and $j$ is denoted by $t$, which is set to $ t = 1$ as the energy unit throughout the paper. The chemical potential $\mu$ and  spin flavor dependent energy $\epsilon_{\alpha}$ together control the population  of each spin component. $U_{\alpha,\beta}$ denote on-site Hubbard repulsive interactions between particles with different spin flavors occupying the same  site. In this work, we use a configuration of $U_{\alpha,\beta}$ breaks the $\mathrm{SU(3)}$ symmetry of the Eq.\ref{eq1}, \textit{i.e.,} we take $U_{\alpha,\beta} = U$ for $[(\alpha,\beta) \in (1,2),(2,3)]$ and $U_{\alpha,\beta} = U^{\prime}$ for $[(\alpha,\beta) = (1,3)]$ and  $U^{\prime} < U$ in general.
For the DMRG calculations,	we consider a one-dimensional chain with  $L-$ sites. The lattice spacing is  set to unity. The open boundary condition is used in our DMRG calculations. We tune $\mu$ and $\epsilon_{\alpha}$  to achieve an equal number of particles of the three spin flavors. The total number of fermions is denoted by $N_e$ , thus the average filling per spin flavor $n = N_e/L$. For example the system is half-filled $n= 1.5$  when $N_e=3L/2$.

	The DMRG approach~\cite{white1992density,schollwock2005density} is a highly accurate numerical method suitable for tackling the ground-state properties of one-dimensional systems. With incorporating  tensor network technique~\cite{evenbly2011tensor}, the computational efficiency of DMRG can be significantly improved. In this method, the variational wavefunction is represented as a matrix product state (MPS), and the many-body Hamiltonian as a matrix product operator (MPO), effectively reducing the Hilbert space dimension to $\chi^2Nd$. Here $\chi$ denotes the MPS bond dimension [i.e., the maximum number of singular values kept in the Singular Value Decomposition (SVD)], and $d$ is the physical dimension of each spin. In this work, we employ the highly efficient tensor network package   $\mathrm{TeNPy}$ ~\cite{tenpy2024} to carry out the computations, where $\chi$ is typically set to $\chi=3000$, with truncation cutoff smaller than $ \epsilon <10^{-6}$.

	\section{Results}
	
	\subsection{Pairing and phase diagram}
	%In both the CDW and superconducting (SC) phases, the spin sector is gapped, and spin correlations decay exponentially. Superconducting properties are typically characterized by the charge density distribution and pairing correlation function.
	Obeying the Mermin-Wagner theorem~\cite{mermin1966absence}, true long-range superconducting order does establish in one-dimensional systems of Eq.~\ref{eq1} at zero temperature. Instead, a quasi-long-range order  similar to the Luther-Emery liquid~\cite{jiang2019superconductivity}, can arise. The system is characterized by hosting algebraically decaying pairing and charge density correlations. The pairing correlation function can be defined as,
	\begin{equation}\label{eq2}
		\Phi(\textit{r})=\langle {\Delta}^\dagger(x_0){\Delta}(x_0+r) \rangle
	\end{equation}
	where  $\Delta_i^{\dagger}=\hat{c}_{i,1}^{\dagger}\hat{c}_{i,3}^{\dagger}$ is the $s-$ wave pairing creation operator of particles with   flavor-1 and   flavor-3 at the same site. In general, $\Phi(r)$ exhibits a power law dependence on $r$,  $\Phi(r)\sim r^{-K_{sc}}$, where $K_{sc}$ is Luttinger exponent for pairing ~\cite{jiang2019superconductivity}.  Note that singlet pairing in one-dimensional chains is restricted to the $s$-wave type~\cite{samokhin2017noncentrosymmetric}. 
	%For the model we study,  we found power laws behavior ofeven very short chains can qualitatively exhibit clear superconducting characteristics. To minimize finite-size effects, a chain of length $L=60$ is used in our calculations.
	In our study, we found that  a chain of 60 sites ($L=60$) is sufficient to reveal the dominant superconducting behavior, where the finite-size effect is negligible.
	Throughout the study we take  $x_0=L/4$ and fit  $\Phi(r) $ data for $r$ in the range of $r= 0\sim L/2$ to obtain Luttinger exponents.

	%When the repulsive Coulomb interactions $U$ and $U^{\prime}$ are absent, quantum fluctuations are very strong, and electrons are fully delocalized in real space, forming a gapless Fermi gas in both charge and spin sectors. Therefore, to investigate the effect of the repulsive interactions on the superconducting properties, we plot in Fig.\ref{fig1} the pairing correlation functions at fixed $U$ while varying $U^{\prime}$.

	\begin{figure}[htbp]
		\centering
		\includegraphics[width=0.35\textwidth]{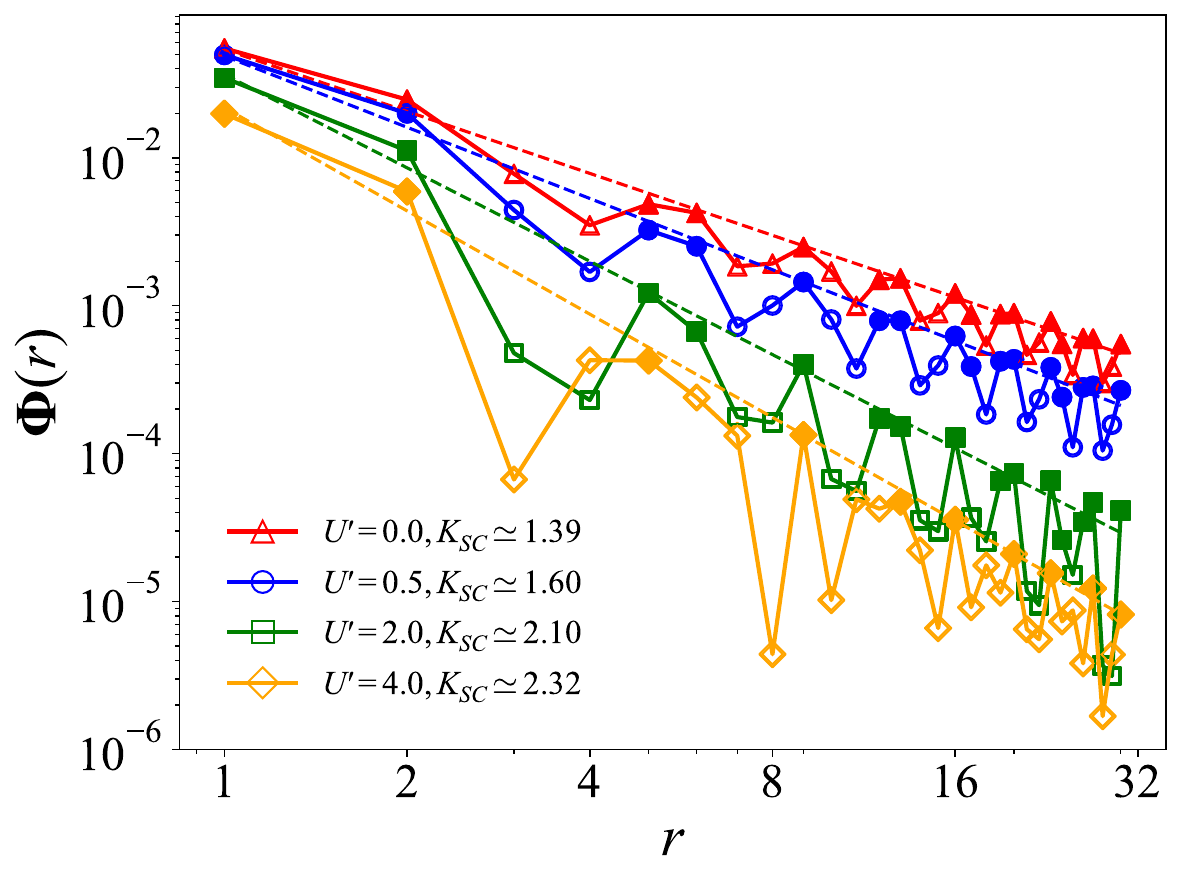}
		\caption{\justifying Pair correlation function \( \Phi(r) \) as a function of $r$ for a few different $U^{\prime} \equiv U_{13}$. Figure is plotted on a double-logarithmic scale. Dashed lines show fitting of \( \Phi(r) \propto r^{-K_{SC}} \). Here \( U \equiv U_{12} \equiv U_{23}= 4.0 \) is fixed and   average filling $n = 0.85$.}
		\label{fig1}
	\end{figure}
	
	In Fig.\ref{fig1}, the pairing correlation function $\Phi(\textit{r})$  at fixed $U=4t$ are plotted as a function of real-space distance $r$ for a few different $U^{\prime}  \equiv U_{13}$. We can see that system  exhibits algebraically decaying $\Phi(\textit{r})$ with  Luttinger exponents $K_{sc} <2$  when $U^{\prime}$ is significantly smaller than $U$ (Triangles and Dots). This result indicates strong and robust superconducting correlations~\cite{peng2025superconductivity} for Cooper pairs comprise of  flavor-1 and  flavor-3 fermions at small  $U^{\prime}$. As  repulsion $U^{\prime}$ increase, Luttinger exponent  $K_{sc}$ increases.  Specifically, when $U^{\prime}\approx U/2$ (Squares in Fig.~\ref{fig1}),   Luttinger exponent $K_{sc}$ approaches two, $K_{sc}\approx2$, representing a critical threshold for the quench of long-range superconducting fluctuations. Indeed, if the pairing susceptibility at finite temperature $T$ can be assumed as  $\chi_{sc}\sim T^{-(2-K_{sc})}$~\cite{arrigoni2004mechanism},  $K_{sc}  \leq 2$ leads to a non-diverging  $\chi_{sc}$ as $T\rightarrow0$. When $U^{\prime}$ is further increased, $U^{\prime} > U/2$,  $\Phi(\textit{r})$  continues to decay algebraically against distance $r$ with $K_{sc}>2$ (Dimonds in Fig.~\ref{fig1}). 
	
\begin{figure}[htbp]
		\centering
		\includegraphics[width=0.35\textwidth]{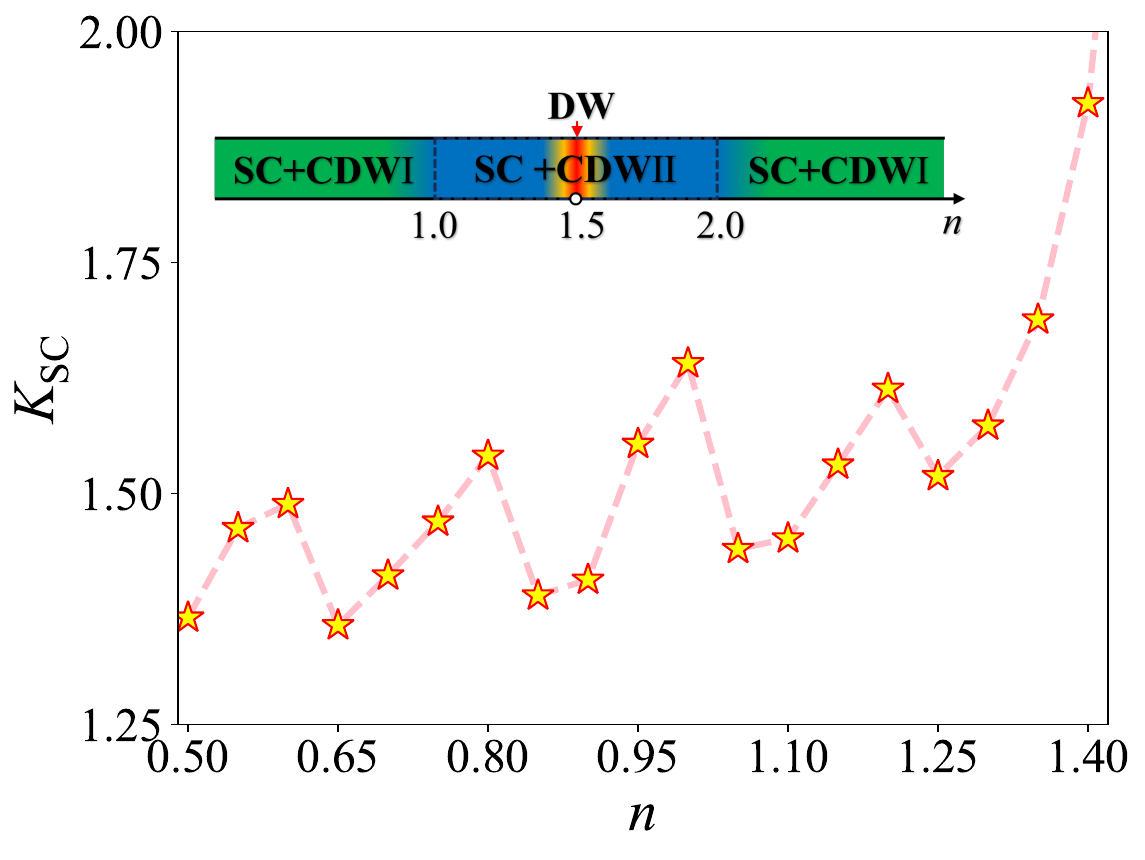}
		\caption{\justifying Luttinger exponent $K_{sc}$ versus particle density $n$ at fixed $U=4.0$ and $U^{\prime}=0$. Inset provides a preliminary delineation of different phase regimes.} 
		\label{fig2}
	\end{figure}

	We now turn to the doping evolution of  superconductivity with  fixed $U=4t$ and $U^{\prime} = 0$.
	%To further investigate the effect of doping concentration on superconducting properties, we consider the case with nearly vanishing $U^{\prime}$. 
	The dependence of the Luttinger exponent $K_{sc}$ on particle density $n$ is shown in Fig.\ref{fig2}. Since there is only nearest-neighbor hopping in our model, superconductivity should be symmetric with respect to hole and electron doping~\cite{grabovsky2019limits}. Without loss of generality, here we present results in the hole-doping regime in Fig.\ref{fig2}. 
	On a general level, Luttinger exponent $K_{sc}$ is less than two $K_{sc} <2$ across the whole doping regime, indicating 
	a universal predominant superconducting fluctuations when systems is doped from half-filling. 
	In Fig.\ref{fig2}, we also notice that the magnitude of Luttinger exponent $K_{sc}$ exhibits an oscillation  with particle density $n$ upon hole doping ($n<1.5$). Notably, $K_{sc}$ reaches local maxima when $n \bmod 0.2 = 0$, where superconducting correlations decay at a faster pace. This is because  stronger  density-wave (DW) fluctuations develop at these fillings, which competes with  pairing.
	
	The overall phase diagram can be roughly divided into several regimes as shown in Inset of Fig.\ref{fig2}. When the total particle density is below 1 (1/3 filling), the charge-density correlation functions exhibit power-law decay, consistent with the hole-doped case in quasi-one-dimensional SU(2) systems~\cite{jiang2024ground}. For fillings with particle density $n$ between $1$ and $1.5$ ( \textit{i.e.,} between 1/3 $\sim$ 1/2 filling), the charge distribution shows a more complex pattern, which can be characterized by a superposition of two  charge density waves with distinct wave vectors. The overall behavior of charge distribution, however, still exhibits a power-law envelope. At half-filling ($n = 1.5$), long-range superconductivity is absent,  the system exhibits  spin flavor dependent density-wave  state, like the antiferromagnetic phase in the $\mathrm{SU(2)}$ Hubbard model, see discussions below.

	\subsection{DW}
\begin{figure}[htbp]
		\centering
		\begin{subfigure}[b]{0.23\textwidth}
			\centering
			\includegraphics[width=\textwidth]{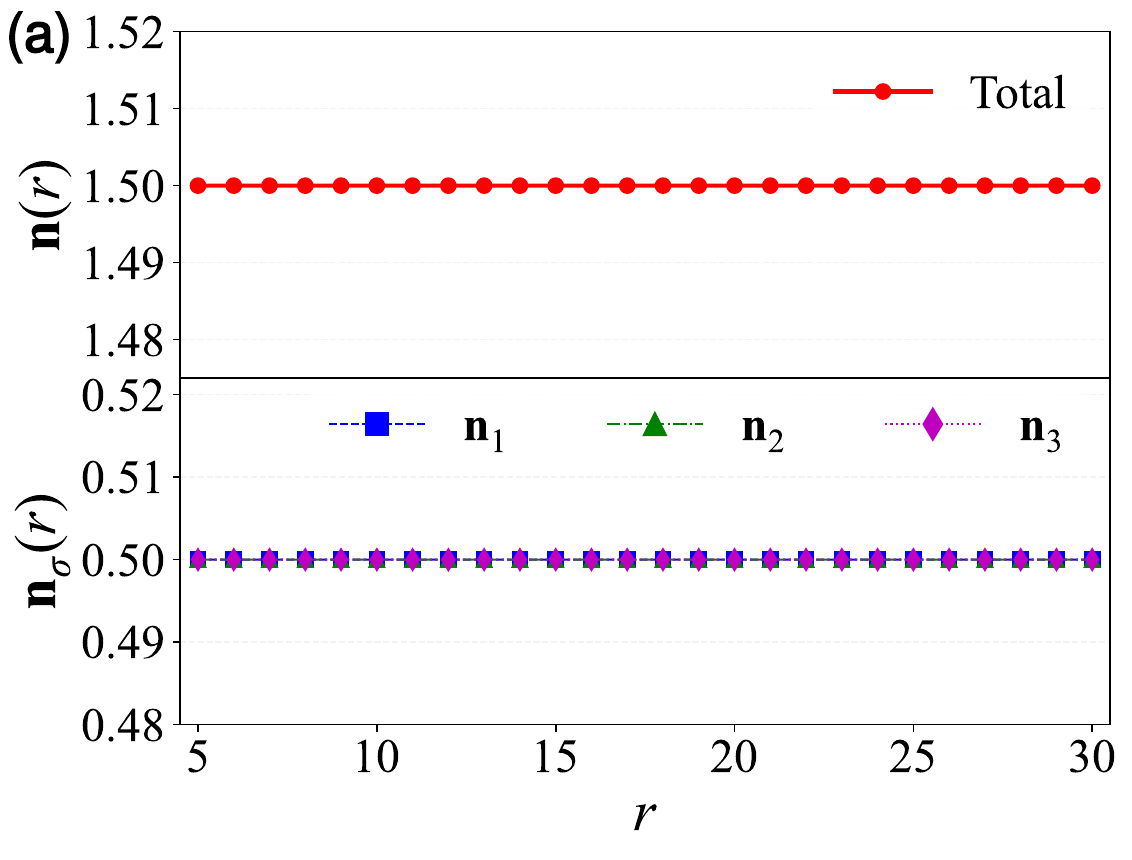}
			\caption{}
			\label{fig3a}
		\end{subfigure}
		\hfill
		\begin{subfigure}[b]{0.23\textwidth}
			\centering
			\includegraphics[width=\textwidth]{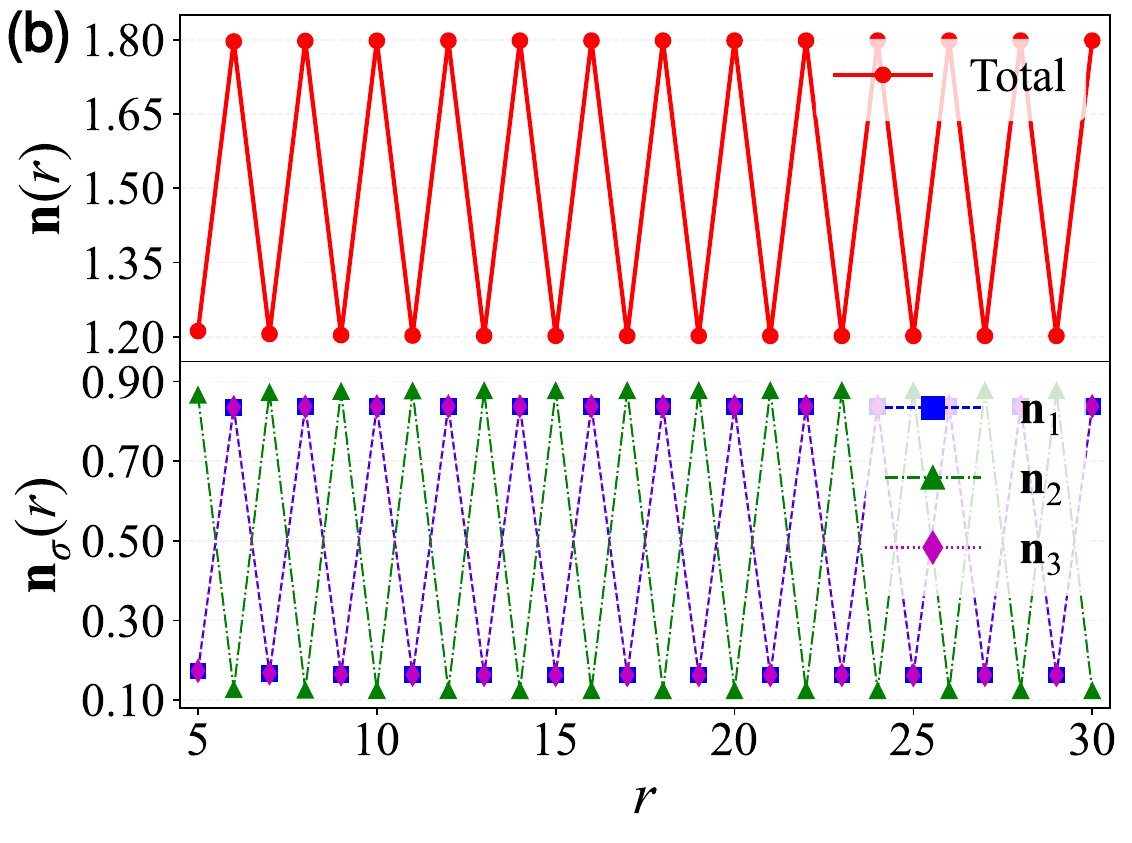}
			\caption{}
			\label{fig3b}
		\end{subfigure}
		\caption{\justifying Aggregated  and spin-flavor resolved charge density distributions at half-filling. (a) For \( U = 2.0 \), (b) For \( U = 4.0 \). Here $U^{\prime} = 0$.}
		\label{fig3ab}
	\end{figure}

	We  now investigate the half-filling case at $N_e = 3L/2$ ($n = 0.5$), where density wave state dominates the low-energy physics. Fig.~\ref{fig3a} and \ref{fig3b} depict the aggregated and spin-resolved charge density distributions at $U=2t$ and $U=4t$, respectively. In the weak-coupling regime ($U=2t$),   three types of particles are more or less evenly distributed on each site with $\langle n_{i\sigma} \rangle = 0.5$, resembling a featureless paramagnetic state. At large $U$ ($U=4t$), pronounced density wave order with a wave length $\lambda_C$ being approximately double lattice spacings, $\lambda_C=2$, emerges, as shown in Fig.~\ref{fig3b}. We have verified that this modulation does not flatten out with increasing the bond dimension~\cite{lu2024emergent}. The $\lambda_C=2$ modulation of $n(r)$ can be understood as, to avoid the potential energy gain, \text{flavor-1} and \text{flavor-3} fermions tend to doubly occupy the same site, while \text{flavor-2} fermions occupy adjacent sites. This state resembles an insulating antiferromagnetic order in the  $\mathrm{SU(2)}$ Hubbard model.
	
	To gain insight, below we  perform analytical investigation into the half-filled  case in strong-coupling limit.	Under the condition of $N_{e1}=N_{e2}=N_{e3}=L/2$,
	as $U\to\infty$,
	the low-energy Hilbert space of our model consists of configurations 
	in which each lattice site hosts either a doublon composed of  flavor-1 and  flavor-3 fermions, or one  flavor-2 fermion. Hence we can define pesudo-spins as
	\begin{align}
		\mid\uparrow\rangle_i\equiv 
		c_{i1}^\dagger c_{i3}^\dagger |\text{vac}\rangle,
		\quad
		\mid\downarrow\rangle_i\equiv 
		c_{i2}^\dagger |\text{vac}\rangle\>.
	\end{align}

	The second-order virtual hopping processes contribute only to the diagonal part of the low-energy effective Hamiltonian. 
	Specifically, a flavor-1 (also flavor-3) fermion can hop from site $i+1$ to $i$ and back, 
	with an intermediate energy cost of $U$. 
	Including both spin flavor components, this process contributes an energy gain of $-2t^2/U$. 
	Similarly, a flavor-2 fermion can hop from $i+1$ to $i$ and back, 
	with an intermediate energy cost of $2U$. 
	Since the flavor-2 fermion has only a single component, 
	this process contribute an energy shift of $-t^2/(2U)$.
	The third-order virtual hopping processes, in contrast, contribute to the off-diagonal part. 
	In this case, flavor-1 and flavor-3 fermions can  move from $i+1$ to $i$, 
	while a flavor-2 fermion simultaneously hops from $i$ to $i+1$. 
	Summing over all $3! = 6$ possible permutations of these processes yields a total energy contribution of $4t^3/U^2$ (see also Appendix).
	
	Along this direction to perform a large-$U$ expansion of Eq.~\ref{eq1} up to third order and dropping the constant shift, the Hamiltonian can be cast into the form of an XXZ chain with
	\begin{align}
		H_\text{eff}=J\sum_i(S_i^x S_{i+1}^x+S_i^y S_{i+1}^y+\Delta S_i^z S_{i+1}^z)
	\end{align}
	where superexchange $J=8t^3/U^2$ 
	and the anisotropy  $\Delta=5U/8t$. The positive $J$ and $\Delta$ ensures AFM order of the pseudo-spins,
	and the stargered magnetization given by Bethe Ansatz,
	\begin{align}
		m_s(\Delta)
		=\frac{1}{2}\prod_{n=1}^{\infty} \tanh^2[n\arccosh(\Delta)]
	\end{align}
	
	For the $U=4t$ case shown in Fig.~\ref{fig3b}, we have $J=1/2$ and $\Delta=5/2$. The infinite product gives $m_s(5/2)=0.416680$.
	Thus
	$
		\langle n_{i,1}\rangle=0.5-(-1)^i (0.5-2*0.416680)		=(0.16664,0.83336,\cdots)
	$.
	The corresponding numerical result from DMRG is found as  (0.1626, 0.8374, $\cdots$ ),  in nice agreement with above analytical predictions. 
	
	%\ww{The total charge density is given by $\langle n_{\mathrm{tot}} \rangle_{\max} = 2\, \langle n_1 \rangle_{\max} + \langle n_2 \rangle_{\min} = 2\, \langle n_1 \rangle_{\min} + \langle n_2 \rangle_{\max}$, so that only in the strong-coupling limit is $\langle n_1 \rangle_{\max} = \langle n_2 \rangle_{\max} = 1.0$.}

\begin{figure}[htbp]
		\centering
		\begin{subfigure}[b]{0.23\textwidth}
			\centering
			\includegraphics[width=\textwidth]{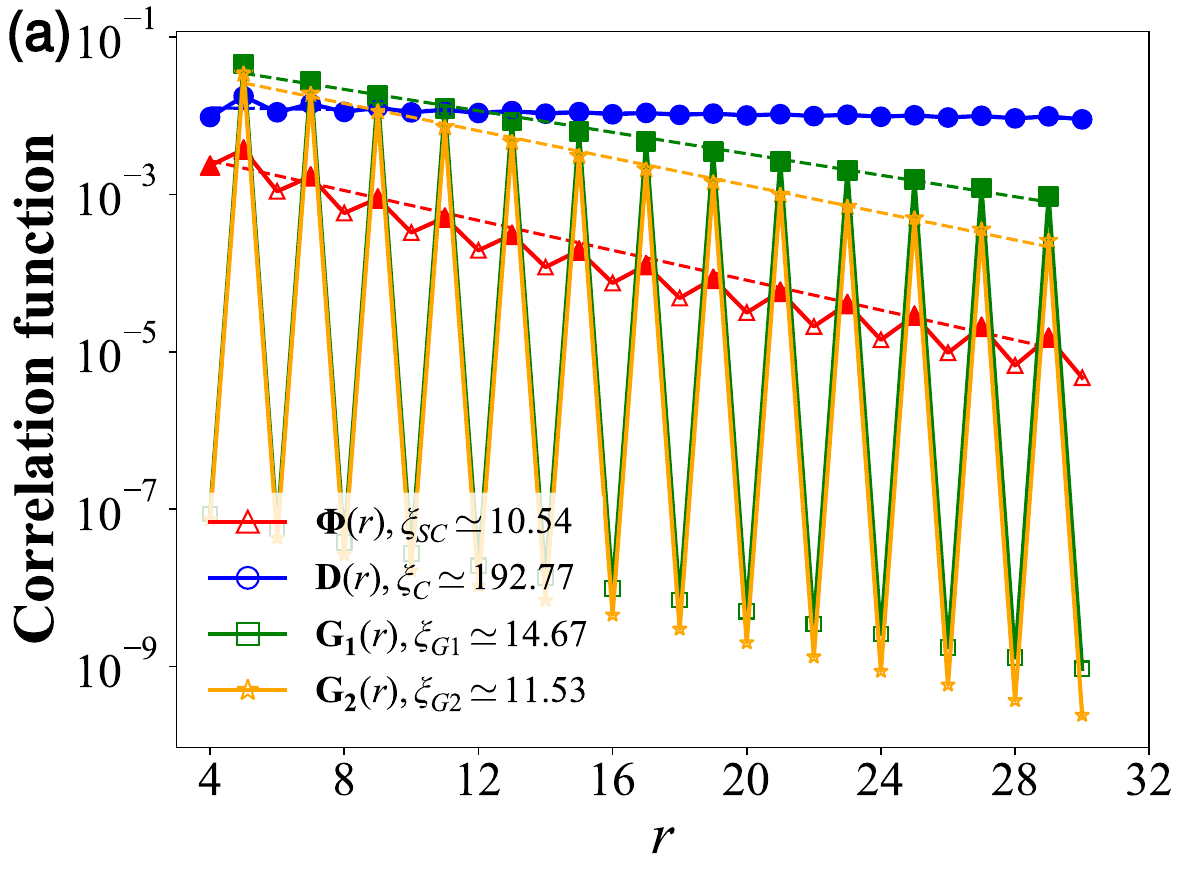}
			\caption{}
			\label{fig4a}
		\end{subfigure}
		\hfill
		\begin{subfigure}[b]{0.23\textwidth}
			\centering
			\includegraphics[width=\textwidth]{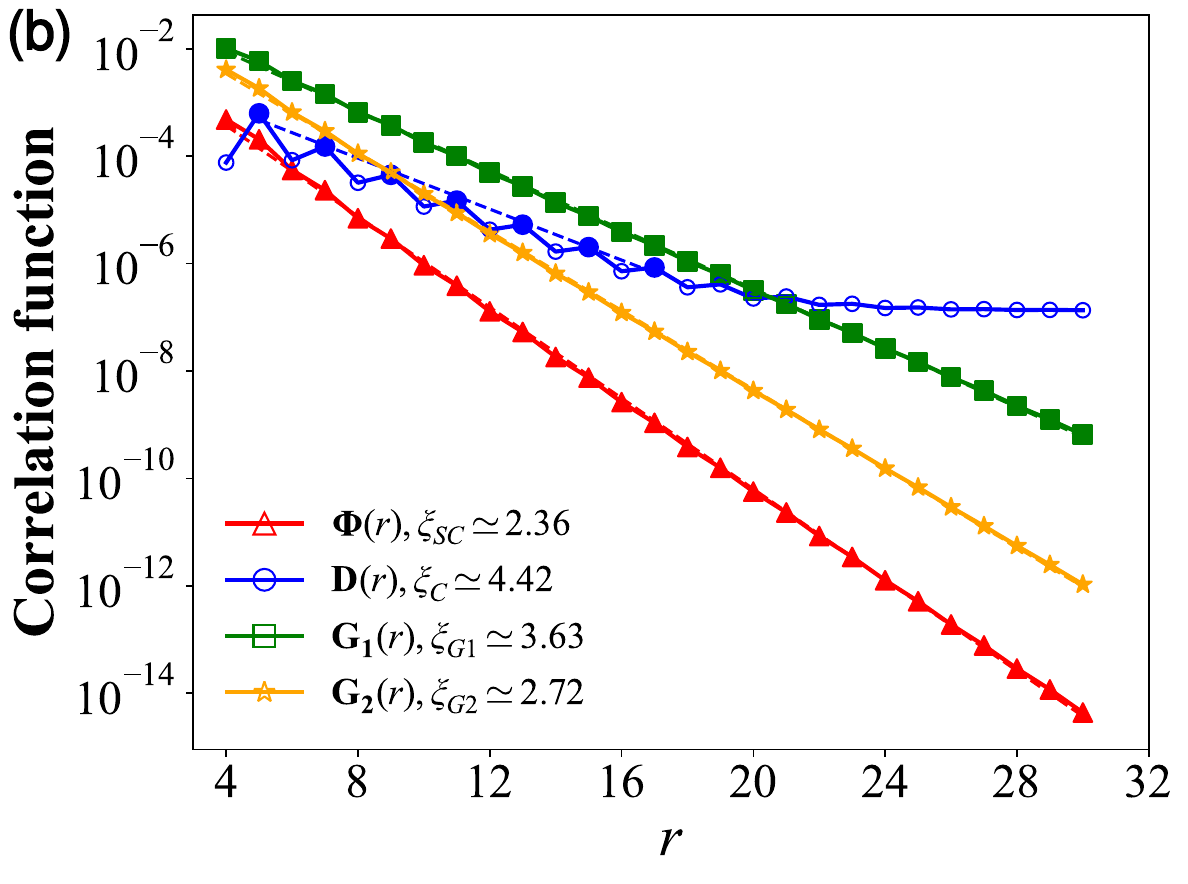}
			\caption{}
			\label{fig4b}
		\end{subfigure}
		\caption{\justifying Correlation functions at half-filling, plotted on a semi-logarithmic scale. (a)For \( U = 2.0 \), (b) For \( U = 4.0 \). Here $U^{\prime} = 0$. Dashed lines show exponential fits. Note that due to symmetry, $G_{1} (r) = G_{3} (r) $. }
		
		\label{fig4ab}
	\end{figure}

	At half-filling, the pairing correlation function decays exponentially $\Phi(r)$ with distance $r$ , indicating no long-range superconducting order, as shown in Fig.~\ref{fig4ab}. To further study the charge and spin dynamics, the charge density-density correlation function $D(r)$ and equal-time single-particle Green’s function $G_{\sigma}(r)$ can be defined as follows:
	\begin{equation}
		D(r) = \left\langle \left( \hat{n}_{x_0} - n_{x_0} \right) \left( \hat{n}_{x_0 + r} - n_{x_0 + r} \right) \right\rangle
	\end{equation}

	\begin{equation}
		G_\sigma (r)=\langle \hat{c}^\dagger_{\sigma,x_0}\hat{c}_{\sigma,x_0+r} \rangle
	\end{equation}
	
	%Fig.\ref{fig4a} and \ref{fig4b} reveal the pronounced sensitivity of the system to the Hubbard interaction parameter $U$ at half-filling, in contrast to the conventional SU(2) system where an infinitesimal $U$ suffices to open a charge gap~\cite{lieb1968absence}.
	
	In Fig.~\ref{fig4a} and Fig.~\ref{fig4b}, we compare $D(r)$ and $G_{\sigma}(r)$ at small and large $U$ respectively.
	One can clearly see that for small $U$ (Fig.\ref{fig4a} ), $D(r)$ remains nearly constant for varying $r$~\cite{feng2023metal},  indicating the enhanced DW fluctuations. When $U$ is increased to $U=4t$ (Fig.\ref{fig4b} ), $D(r)$ decays exponentially at small $r$, reflecting the localization of particles as a result of large  $U$, which suppresses DW fluctuations. Remarkably, at long distance,$D(r)$  does not decay with $r$, despite its magnitude is diminished, see Fig.\ref{fig4b}. This may due to the fact that our model has three components of fermions with uneven mutual repulsions $U \neq U^{\prime}$. Degenerate DW configurations in particle distribution are favored from the energetic aspect.
	
	From Fig.~\ref{fig4a} and Fig.~\ref{fig4b} we can see that both $G_1(r)$ [ also $G_3(r)$ ] and $G_2(r)$ decay exponentially for small and large $U$, indicating the absence of  low-energy single-particle excitations, like in the conventional SU(2) system where an infinitesimal $U$ suffices to open a charge gap~\cite{lieb1968absence,schlomer2024subdimensional} at half-filling. 
	%Notably, we observe that correlation lengths $\xi_{G2}\approx\xi_{SC}$ for different $U$, both exhibiting relatively short correlation lengths. This suggests that short-range superconducting correlations may be closely linked to the single particle excitation of  \text{spin flavor-2} particles.
	% Furthermore, since $\xi_{G1}>\xi_{G2}$ and the spin gap $\Delta \sim 1/\xi_G$ , \text{flavor-1} excitations require relatively low energy. \ww{This may be attributed to the tendency of \text{flavor-1} states to form \text{flavor-3} pairs, significantly affecting neighboring \text{flavor-1} and \text{flavor-3} states and leading to extended correlations.}

	\subsection{SC+CDWI}
	Upon doing, the dominant density wave state at half-filling becomes suppressed and long-ranged superconducting correlations develop in the system when $U^{\prime} <<U$, leading to the coexistence of  charge density wave and long-ranged superconducting fluctuations.
	%The charge density wave that coexists with the superconducting fluctuations.
	The local charge density distribution $n(x)$ follows the  Friedel oscillation relation~\cite{white2002friedel}  introduced by the open boundaries,
	\begin{equation}\label{eq3}
		n(r)=n_0+\delta n cos(2k_Fr+\phi)r^{-K_c/2}
	\end{equation}
	where $n(r)$ is the density summed over spin flavors $n(r) = \sum_{\alpha} n_{r,\alpha}$. The Luttinger exponent $K_c$ also characterizes the power-law decay of the charge density-density correlations~\cite{jiang2019superconductivity}, which can be obtained by fitting our numerical data of $n(r)$ with Eq.~\ref{eq3}.   There are also a few other fitting parameters: $\delta n$ is a non-universal amplitude, and $n_0$ denotes the value of the background density. $\phi$ is the phase shift, and $k_F$ is the Fermi wave vector.
	
\begin{figure}[htbp]
		\centering
		\includegraphics[width=0.35\textwidth]{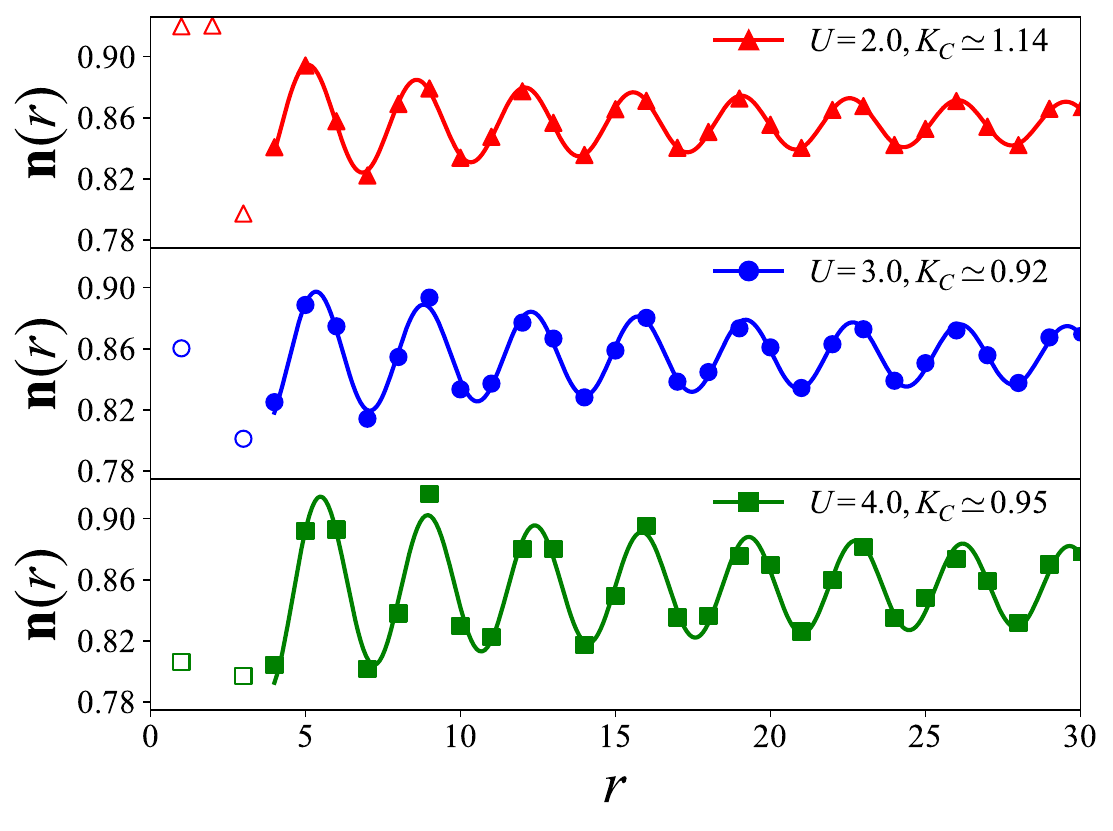}
		\caption{\justifying Charge density distribution $n(r)$ for different $U$. Solid lines showing fitting curves obtained using Eq.~\eqref{eq3}. Here $n=0.85, U^{\prime} = 0$.}
		\label{fig5}
	\end{figure}

	Fig.\ref{fig5} illustrates the charge density distributions  $n(r)$ for different interaction strengths $U$, in the over-doing regime [$N_e \in (0, L), n\in (0,1/3)$]. Excluding a few boundary sites, one can clearly see that the charge distributions can be well captured by the Friedel oscillation ( solid lines) described by Eq.~\eqref{eq3}. As $U$ increases, the system exhibits more complex behaviour: the CDW correlations become enhanced (as indicated by a decreasing $K_c$), while $n(r)$ becomes irregular, starting to form a quasi-periodic pattern that is not strictly oscillating according to Eq.~\eqref{eq3}.
	 %\ww{In particular, \text{spin-0} and \text{spin-2} components show stronger local occupation in certain adjacent regimes}, giving rise to a long-range ordered sublattice structure. This may result from a strong short-range pairing between nearby lattice sites \ww{[more reasoning on this point] }.
	Similar to the half-filling case, here we also find that the occupation numbers of flavor-1 and flavor-3 components show cooperative enhancement/suppression  on adjacent sites , accompanied by a corresponding suppression / enhancement of the flavor-2 component (data not shown). Obviously, such an occupation pattern is again due to the fact that $U>U^{\prime}$ , which establishes a long-range quasi-sublattice structure throughout the system in doped cases.
	
\begin{figure}[htbp]
		\centering
		\includegraphics[width=0.35\textwidth]{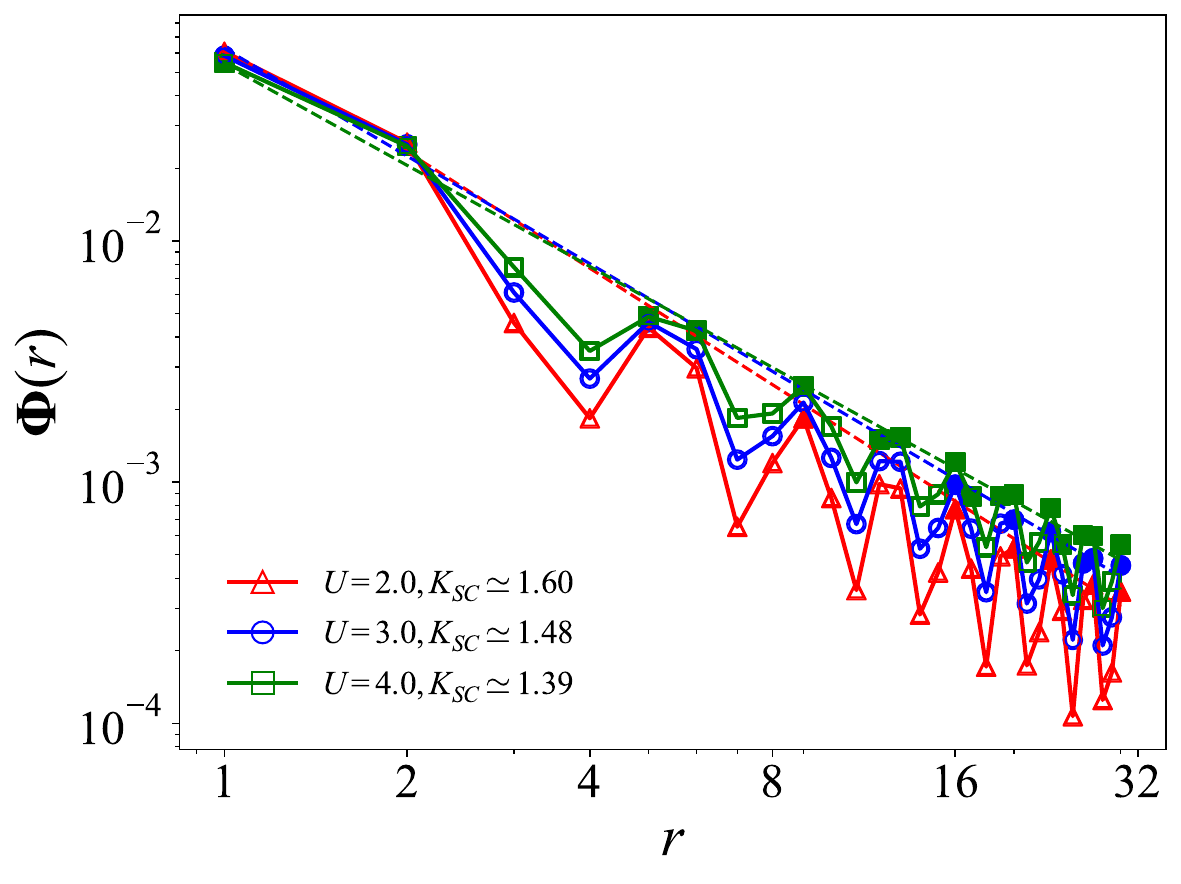}
		\caption{\justifying Pairing correlation \( \Phi(r) \sim r^{-K_{\mathrm{SC}}} \)  as a function of $r$ for a few different $U$. 
		The figure is plotted on a double-logarithmic scale at $n = 0.85, U^{\prime} = 0$.}
		\label{fig6}
	\end{figure}

	Fig.\ref{fig6} displays pairing correlation functions in this SC+CDW {I} phase, where algebraic decay is observed over a wide range of interaction strengths $U$~\cite{lu2023ground}. 
	%\ww{Here both the pair correlation function $\Phi(r)$ and the charge density $n(x)$ exhibit scaling laws characteristic of the charge sector~\cite{lu2023ground}}. 
	Due to the presence of CDW modulations, $\Phi(r)$ shows spatial oscillations similar to those of $n(r)$,  whose oscillation period exactly matches the charge density wavelength $\pi/k_{F} = 2/{(1-\delta)} =3L/N_e$. As on-site repulsion $U$ increases, Luttinger exponent ${K_{sc}}$ gradually decreases, reflecting the enhancement of superconductivity with $U$. %\ww{ However, pairing correlations alone cannot fully establish whether the system forms a true superconducting phase}. %For the LE liquid state, the product ${K_c \times K_{sc}}$ is expected to be close to unity~\cite{jiang2018superconductivity}. This relation holds reasonably well for sufficiently large $U$, while for weak interactions, ${K_C\times K_{SC}}$ deviates significantly from unity, indicating that the system is closer to a Luttinger liquid regime \ww{data/evidence?}.
	%However, the power-law decay of the pairing correlations alone cannot fully establish whether the system forms a true superconducting phase. 
	 We find that the relation ${K_c\times K_{sc} \simeq 1.0}$ in general holds for sufficiently large $U$ (e.g., at $U=4.0$, we find $K_c \times K_{sc} \simeq 1.3$, consistent with the expectation for a Luther--Emery (LE) liquid~\cite{jiang2019superconductivity} within numerical uncertainties . For weaker interactions, the product ${K_c\times K_{sc}}$ deviates significantly from unity(e.g., at $U=2.0$, $K_c \times K_{sc} \simeq 1.8$)), indicating that the system moves closer to a Luttinger--liquid regime.

\begin{figure}[htbp]
		\centering
		\begin{subfigure}[b]{0.23\textwidth}
			\centering
			\includegraphics[width=\textwidth]{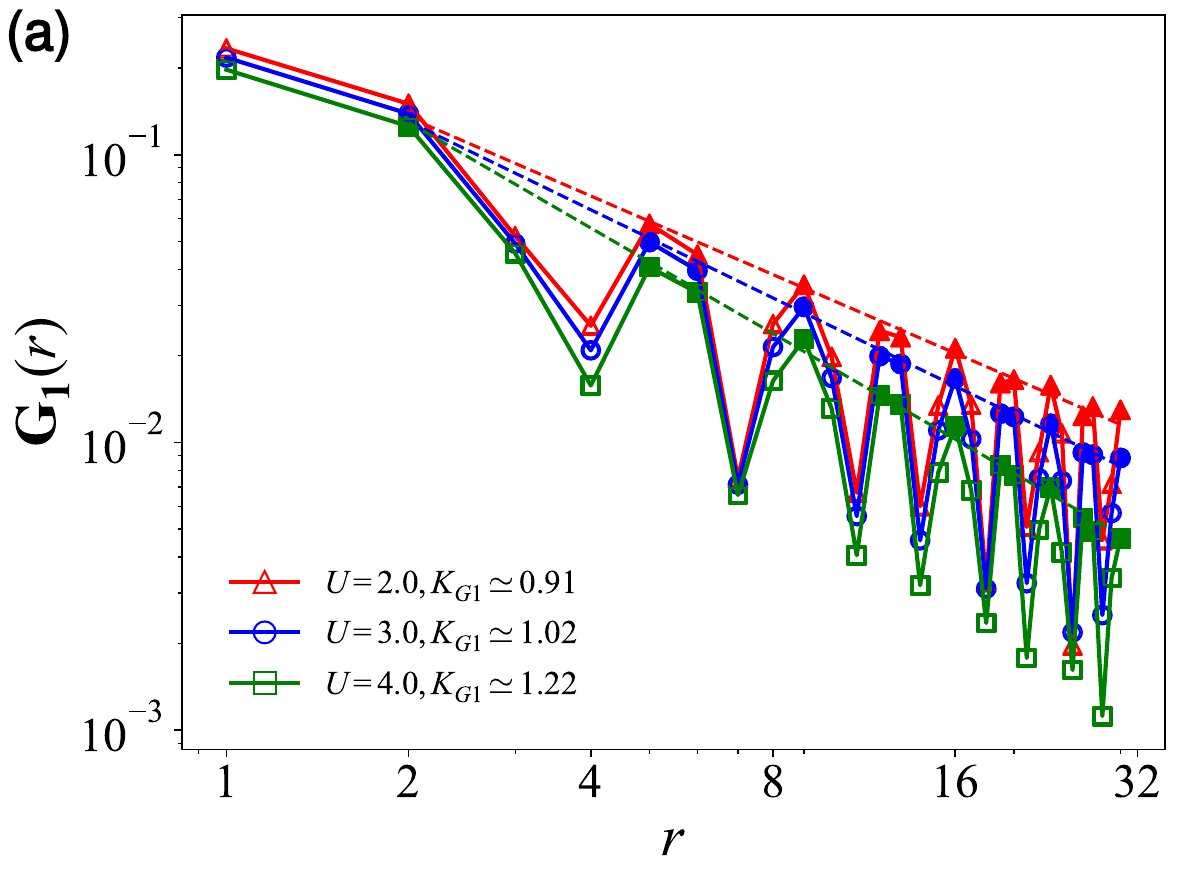}
			\caption{}
			\label{fig7a}
		\end{subfigure}
		\hfill
		\begin{subfigure}[b]{0.23\textwidth}
			\centering
			\includegraphics[width=\textwidth]{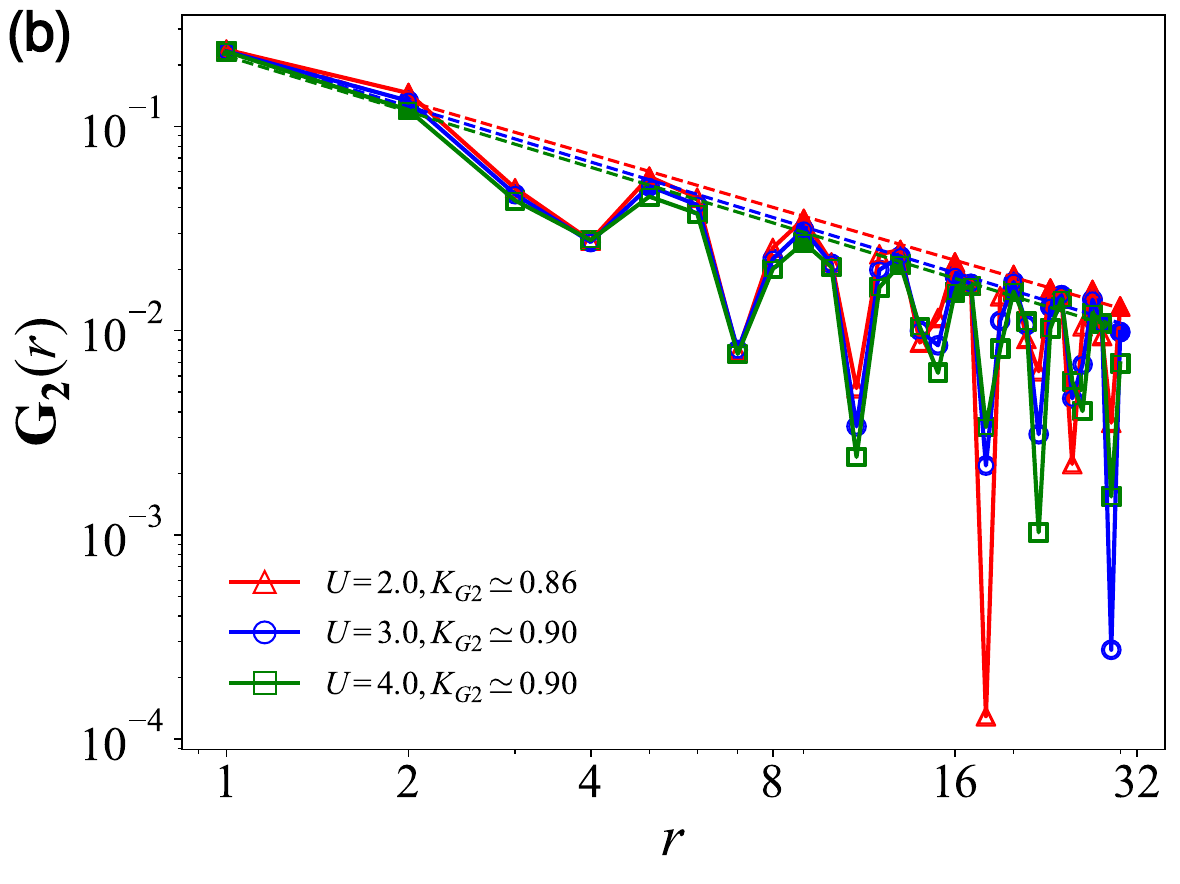}
			\caption{}
			\label{fig7b}
		\end{subfigure}
		\caption{\justifying Single-particle correlation function $G_{\sigma} (r)$ as a function of $r$ for different $U$, plotted on a double-logarithmic scale. (a) For  $G_1(r)$. Dashed lines show fitting of \( G_1(r) \sim r^{-K_{G1}} \). (b) For  $G_2(r)$. Dashed lines show fitting of \( G_2(r) \sim r^{-K_{G2}} \). Here $n=0.85$.}
	
		\label{fig:both}
	\end{figure}

	The single-particle correlation functions for \text{flavor-1} ( also \text{flavor-3}) and \text{flavor-2} components both exhibit power-law decay in  the SC+CDW {I} regime, indicating the presence of single particle excitations, as shown in Fig.~\ref{fig7a} and Fig.~\ref{fig7b}. When Coulomb repulsion $U$ increases, single particle Luttinger exponents $K_{G1}$($K_{G3}$) grows rapidly, signaling a suppression of long-range itinerancy for \text{flavor-1} and \text{flavor-3} particles. This is due to the rising of long-range superconducting fluctuation, where  flavors-1 and flavor-3 fermions stars to bind, leading to collective Cooper pair excitations.
It is worthy noting that here we find $K_{\rm sc} < K_{G1} + K_{G3}$,  which also suggests the presence of  preformed Cooper pairs, since $\Phi(r)$ decays more slowly than $|G_1(r)|*|G_3(r)|$.	
	On the other hand, the \text{flavor-2} component, which mediates the pairing, remains comparatively free to move since its Luttinger exponent $K_{G2}$ are significantly smaller than $K_{G1}$($K_{G3}$), see Fig.~\ref{fig7b}.
	\subsection{SC+CDWII}
\begin{figure}[htbp]
		\centering
		\includegraphics[width=0.35\textwidth]{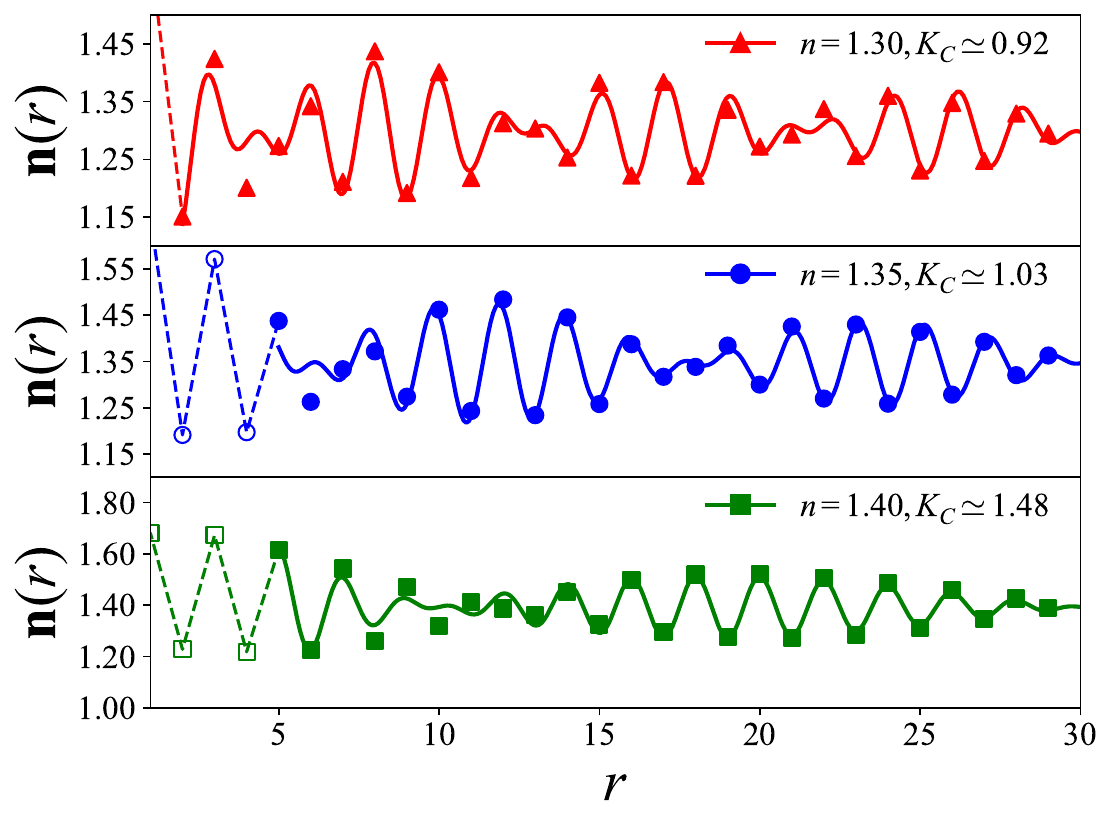}
		\caption{\justifying Charge density distribution $n(r)$ at \( U = 4.0 \) as a function of $r$ at different particle densities $n$. Solid lines showing fitting of the data with Eq.~\eqref{eq5}. The overall shapes of $n(r) $ resemble a series of wave packets, with boundaries sites containing approximately half of a wave packet.  }
		\label{fig8}
	\end{figure}

	Comparing to the over-doped cases, when particle number per site $n$ is between 1 and 1.5 (half-filling),  charge density distribution $n(r)$ show more complicated behaviours, which can be characterized as the superposition of two CDWs with different wavelengths. The two waves share the same phase and amplitude, manifesting as symmetric beating wave packets with half-packets at the boundaries (Fig.~\ref{fig8}). This state can be interpreted as the beating of multiple CDWs, or a kinked CDW~\cite{zhong2024density}. One therefore can describe the density distribution as a superposition of two Friedel oscillations sharing a common Luttinger exponent $K_c$:
	\begin{align}\label{eq5}
		n(r)=&n_0+\delta n cos(2k_{F1}r+\phi)r^{-K_C/2}\notag\\
		&+\delta n cos(2k_{F2}r+\phi)r^{-K_C/2}
	\end{align}
	the wave vectors $k_{F1}$ and $k_{F2}$ of the two charge density waves are related to the number of wave packets $n_{wp}$ by: 
	\begin{equation}
		k_{F1}-k_{F2}=\pi n_{wp}/L
	\end{equation}
	
	In Fig.~\ref{fig8}, $n(r)$ is plotted as a function of $r$ for three different $n$ in the under-doped [$n \in (1, 1.5)$] regime. We can see that while the wave packet of $n(r)$ exhibits oscillations,  the overall envelope of $n(r)$ decays with $r$, leading to the gradual reduction in peak amplitudes. 
	In Fig.~\ref{fig8} the fitted Luttinger exponent $K_c$ are labelled in the plots, which in general increases with particle density $n$. Note that here the constraint $ 1/2<K_{c}<2$ is imposed in the fitting procedure, as to ensure the Lorentz symmetry of ${K_{sc}}$~\cite{jiang2018superconductivity}.
	% Note that this is only a qualitative result, as the fitting is highly sensitive to boundary points.
	It is interesting to note that the number of charge density wave pockets increase with doping level in our study. For example, at half-filling ($n = 1.5$), the system exhibits a pure charge density wave without any wave packets (see Fig.~\ref{fig3b}). For $n = 1.40$, $1.35$, and $1.30$, there are 3, 5, and 7 wave packets can be identified in Fig.~\ref{fig8}, respectively.
	
\begin{figure}[htbp]
		\centering
		\includegraphics[width=0.35\textwidth]{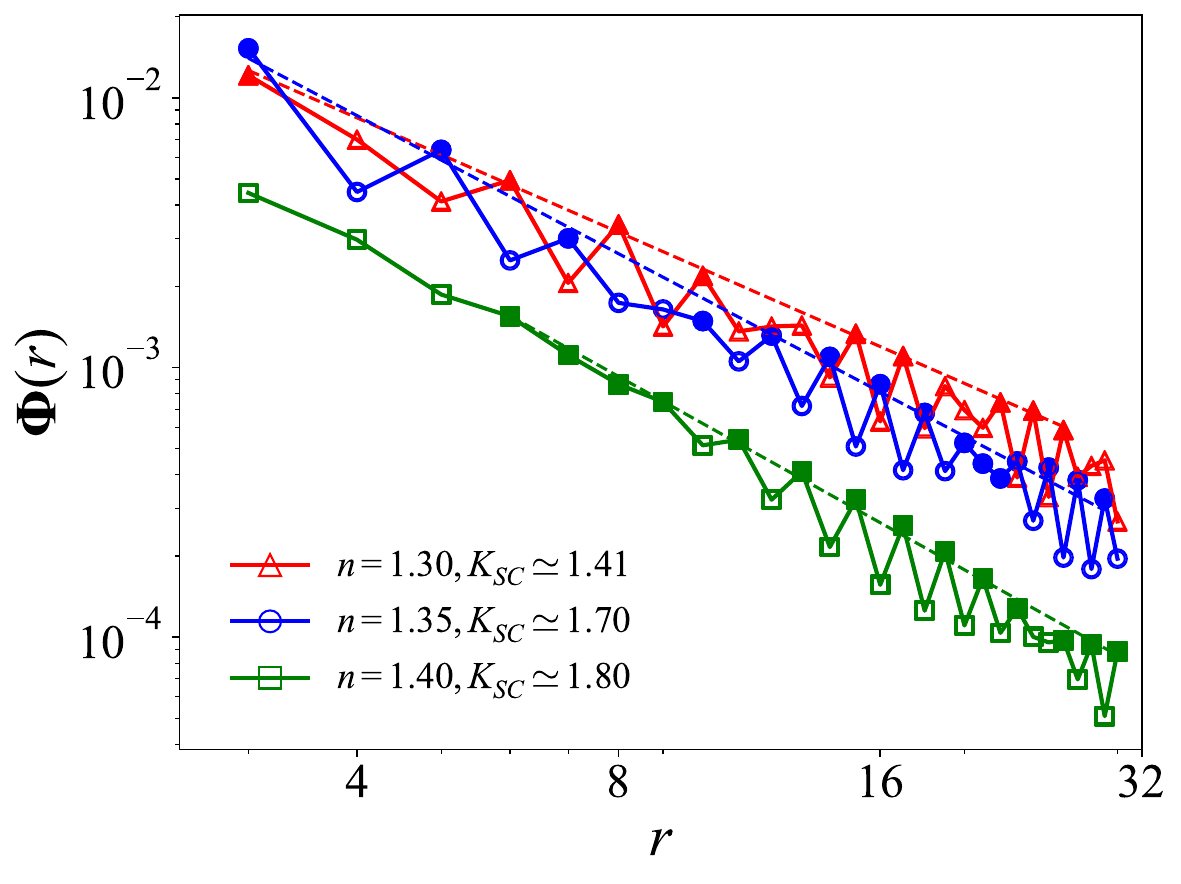}
		\caption{\justifying 
			Pairing correlation \( \Phi(r) \sim r^{-K_{\mathrm{SC}}} \) for three different $N_e$ plotted on a double-logarithmic scale. Here $U=4t.$}
		\label{fig9}
	\end{figure}

	Fig.\ref{fig9} depicts the pairing correlation functions $\Phi(r)$ in the SC+CDW{II} phase for a few different $n$. As one can see, $\Phi(r)$ exhibit strong irregular oscillations while power-law decay is still overall obeyed in this regime. When  particle number $n$ approaches half-filling, $n=1.5$, the Luttinger exponent $K_{sc}$ approaches 2.0,  leading to the rapidly decaying superconducting fluctuations. Examining the pairing correlation decay at $n=1.40$, one can see that it is dominated by a power-law behavior at small $r$  with suppressed oscillations, indicating the prevailing of superconducting correlations. At large $r$, $\Phi(r)$ oscillates with a wave vector $q=2k_F\sim \pi$. 
	
\begin{figure}[htbp]
		\centering
		\includegraphics[width=0.35\textwidth]{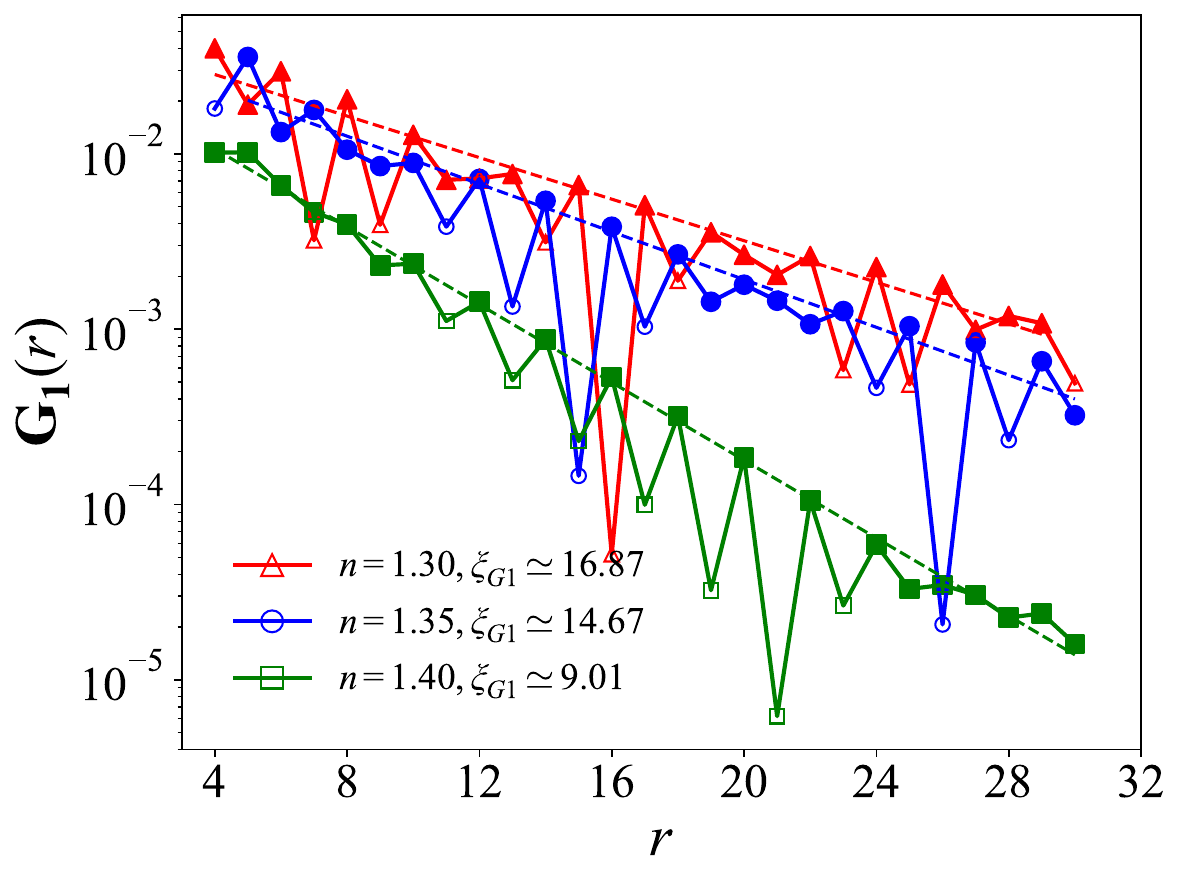}
		\caption{\justifying Single-particle correlation $G_1(r) \sim e^{-r/\xi_{G_1}}$, plotted on a semi-logarithmic scale.}
		\label{fig10}
	\end{figure}

	In Fig.\ref{fig10}, the single-particle correlation function $G_1(r)$ for flavor-1 component is plotted in the  SC+CDW{II} regime, which exhibits nearly exponentially decay with $r$ rather than a algebraical one, in contrast to the case in the SC+CDW{I} regime. This result suggest that the system hosts physical states with fully different  single-particle properties for $n < 1$ and $n > 1$~\cite{jiang2020ground}. Although the correlation length of $G_1(r)$ is relatively long, it remains limited compared to the lattice size. This result suggests that in the  SC+CDW{II} regime  the single-particle excitations of \text{flavor-1} and \text{flavor-3} are quenched in low-energy physics. The system is more likely characterized by collective excitations in the form of Cooper pairs.
	
	\section{Discussion  and conclusion}
	Based on the DMRG method, we investigates the ground-state properties of an one-dimensional three component Hubbard model with an asymmetric Hubbard interaction term. We find that within a wide range of doping level and repulsive interaction strengths, the system exhibits dominant superconductivity long-range fluctuations which coexists with charge density waves. This observation contrasts the conventional two-component nearest-neighbor hopping repulsive Hubbard model, where long-ranged dominant superconductivity fluctuations are absent. Our calculations reveal rich low-energy physics in this system, such as the transition between a single-wavevector CDW {I} state to a superposition of multiple-wavevector CDW {II} state under doping.
	% These results collectively demonstrate the crucial role of SU(3) symmetry breaking in inducing unconventional pairing in repulsive systems.
From an analytic perspective,  both our weak-coupling effective theory and bosonization treatment indicate the emergence of superconductivity in this model (see Appendix), consistent with 
the DMRG result.In particular, we find that pairing between flavor-1 and flavor-3 fermions is mediated by particle-hole fluctuations of the flavor-2 fermions. This mechanism realizes an unconventional form of superconductivity linked to excitonic effects~\cite{little1964possibility,davis1976proposed}.  In essence, our work demonstrates that extending the SU(2) Hubbard model to an SU(3) model with asymmetric interactions offers a novel perspective for exploring unconventional superconductivity and other strongly correlated states. Although our study focuses on one-dimensional systems, the underlying physical mechanism is expected to extend to higher-dimensional lattices.

Finally, we note that the SU($N$) Hubbard model, with $N$ as high as 10, can be controllably realized in ultra-cold atoms within optical lattices~\cite{capponi2016phases,honerkamp2004ultracold,hermele2009mott,milner2025coherent}. This experimental platform may be utilized for future verification of the superconductivity predicted in our three-component ($N=3$) Hubbard model.

\section{Acknowledgment-} This work is supported by the National Natural Science Foundation of China (Grants No.12274472). We also thank the support from the Research Center for Magnetoelectric Physics of Guangdong Province (Grants No. 2024B0303390001).

\appendix

\section{Weak coupling Effective theory}
Below we present a weak-coupling effective theory analysis of the our Hamiltonian to reveal the effective attraction between spin flavor 1 and spin flavor 3 fermions. First, we slightly modify the spin flavor index as: 
\begin{align}
	2\to0, \qquad 1\to\downarrow, \qquad 3\to\uparrow \>.
\end{align}
The Hamiltonian is then decomposed 
according to whether the terms contain the 0-fermion,
\begin{align}
	&\mathcal{H}=\mathcal{H}_{\uparrow\downarrow}+\mathcal{H}_0 \>,\\
	&\mathcal{H}_{\uparrow\downarrow}
	=-t\sum_{i,\sigma}
	(c_{i,\sigma}^\dagger c_{i+1,\sigma}+h.c.)
	+U\sum_{i} n_{i\uparrow} n_{i\downarrow} \>,\\
	&\mathcal{H}_{0}
	=-t_0\sum_{i}
	(c_{i,0}^\dagger c_{i+1,0}+h.c.)
	+U_0\sum_{i} n_{i,0} (n_{i\uparrow} + n_{i\downarrow})\>.
\end{align}

Applying the Jordan-Wigner transformation,
the spinless fermions are mapped to Pauli matrices,
\begin{align}
	c_{i,0}
	=(-1)^{N}\Big(\prod_{j<i}\sigma_j^z\Big)\sigma_i^-\>,\quad
	c_{i,0}^\dagger 
	=(-1)^{N}\sigma_i^\dagger \Big(\prod_{j<i}\sigma_j^z\Big)\>,
\end{align}
where $N=\sum_{i,\sigma} c_{i\sigma}^\dagger c_{i\sigma}$.
The Klein factor $(-1)^N$ ensures the correct
anticommutation between different flavors.

This yields the equavilent Hamiltonian
\begin{align}
	\mathcal{H}
	=&-t\sum_{i,\sigma}
	(c_{i,\sigma}^\dagger c_{i+1,\sigma}+h.c.)
	-\mu \sum_{i} n_i
	+\frac{U}{2}\sum_{i} n_{i}^2 \nonumber\\
	&
	+g \sum_{i}\sigma_{i}^z n_{i}
	+J\sum_i (\sigma_i^x \sigma_{i+1}^x+\sigma_i^y \sigma_{i+1}^y)
	-h\sum_i \sigma_{i}^z \>,
\end{align}
where $n_i=n_{i\uparrow}+n_{i\downarrow}$.
Neglecting the Hubbard $U$, 
the model reduces to a one-dimensional version of spin-fermion model.

rewrite $c_0=\psi$, the full action
\begin{align}
	S[c_{i\sigma},\psi]
	&= \int^{\infty} d\tau \Big[
	\sum_{i\sigma} c_{i\sigma}^\dagger (\partial_\tau - \mu) c_{i\sigma} \nonumber \\
	&\quad + \sum_i \psi_i^\dagger (\partial_\tau - \mu_0) \psi_i
	+ H_{\uparrow\downarrow}[c_\sigma] + H_0[\psi]
	\Big]
\end{align}
the partition function
\begin{align}
	Z=\int D[c_{\sigma},\psi] e^{-S}
	=\int D[c_{\sigma}]e^{-S_\text{eff}[c_{\sigma}]}
\end{align}
Since $H_0[\psi]$ is qudartic in $\psi$,
performing the Gaussian intgeral lead to
\begin{align}
	S_\text{eff}=\sum_{i\sigma} c_{i\sigma}^\dagger (\partial_\tau -\mu) c_{i\sigma}
	+H_{\uparrow\downarrow}[c_{\sigma}]
	+\tr\ln(G_0^{-1}+Un)
\end{align}

Due to the weak coupling expansion,
\begin{align}
	\tr\ln(G_0^{-1}+U_0\rho)
	=\tr\ln G_0^{-1}+\tr\ln(1+G_0U\rho) \nonumber \\
	=\tr\ln G_0^{-1}+\tr(G_0Un)-\frac{1}{2}\tr[(G_0Un)^2]+\mathcal{O}(n^3)
\end{align}
the effective action takes the form
\begin{align}
	S_{eff}=&S_{\uparrow\downarrow}
	+U \langle n_0\rangle\int d\tau\sum_i\rho_i(\tau)\nonumber\\
	&+\frac{U}{2}\int \rho(-q,-\omega)\chi_0(q,\omega)\rho(q,\omega)
	+\cdots
\end{align}

In the static limit $\omega\to0$,
one reach the effective Hamiltonian
\begin{align}
	H_\text{eff}
	=\mathcal{H}_{\uparrow\downarrow}
	+U \langle n_0\rangle \sum_i n_i
	+\frac{U^2}{2}\sum_{q}n_{-q}\chi_{0}(q)n_{q}
\end{align}
where the static polarization is defined as
\begin{align}
	\chi_{0}(q)
	=\sum_{k}
	\frac{n_{0}(k)-n_{0}(k+q)}
	{\epsilon_{k} - \epsilon_{k+q} + i0^{+}} \>,
\end{align}
and $n_{0}(k)$ is the occupation number of 0-fermion.
At the zero temperature,
\begin{align}
	\chi_{0}(q)
	&=\frac{1}{2t}\int \frac{dk}{2\pi}
	\frac{n_{0}(k)-n_{0}(k+q)} 
	{\cos(k)-\cos(k+q)}\nonumber\\
	&=-\frac{1}{4t\pi}\frac{1}{\sin(q/2)}
	\ln\left|\frac{\tan(k_F/2+q/4)}{\tan(k_F/2-q/4)}\right|\>,	    
\end{align}
where the Fermi wave-vector $k_F=\arccos\frac{-\mu}{2t}$.

The polarization takes the value
\begin{align}
	\chi_{0}(q=0)
	=\frac{1}{2\pi t \sin k_F}
	=-\frac{1}{\pi\sqrt{4t^2-\mu^2}}
\end{align}
and
\begin{align}
	\chi_{0}(q=2k_F+\delta q)
	=\frac{1}{4t\pi}\frac{1}{\sin k_F}\ln|\delta q|
	=\frac{\ln|\delta q|}{2\pi\sqrt{4t^2-\mu^2}}
\end{align}

Applying the inverse Fourier transformation from momentum space to real space leads to

\begin{align}
	\chi_0(r) 
	= \sum_{m=0}^{r-1} \frac{\sin[(2m+1) k_F]}{\pi (2m+1)} - \frac{1}{4}, 
	\quad r \in \mathbb{Z}
\end{align}
The onsite term is independent of $k_F$, and the long distance show Friedel oscillation. 

Thus the effective Hamiltonian can be cast into real space as
\begin{align}
	H_\text{eff}
	=&-t\sum_{i,\sigma}
	(c_{i,\sigma}^\dagger c_{i+1,\sigma}+h.c.)\nonumber\\
	&+\frac{U'}{2}\sum_{i} n_{i}^2 
	+\frac{U^2}{2}\sum_{i,r} \chi_{0}(r) n_i  n_{i+r}
\end{align}

When $U' < k U^2$ ($k>0$), the interaction between spin flavor-1 and flavor-3 becomes attractive, leading to the emergence of SC order.

\section{Luttinger Parameters from Bosonization}

\subsection{Bosonization of the Three-Flavor Fermion Model}

Considering the filling $\nu_\sigma$ for each flavor, the Fermi wavevector is given by $k_{F\sigma} = \pi \nu_\sigma$, and the single-particle dispersion
$\epsilon_k = -2t \cos k$ leads to a Fermi velocity $v_F = 2t \sin k_{F\sigma}$.

Below, we assume equal Fermi velocities for all three flavors. The kinetic term is given by
\begin{align}
	H_{0,\text{kin}}
	=\sum_{\sigma}\frac{v_{F,\sigma}}{2\pi}\!\int_0^L\!dx\;
	\big[(\partial_x\phi_{\sigma})^2+(\partial_x\theta_{\sigma})^2\big]
	+\mathcal O(L^{-1}),
\end{align}
with
\begin{align}
	c_{\sigma}(x)\to e^{+ik_{F,\sigma}x}\psi_{R,\sigma}(x)
	+e^{-ik_{F,\sigma}x}\psi_{L,\sigma}(x),
\end{align}
and standard bosonization:
\begin{align}
	&\psi_{r,a}(x)=\frac{U_{r,a}}{\sqrt{2\pi\alpha}}\,
	e^{-i\,(r\,\phi_a-\theta_a)}, \nonumber\\
	&[\phi_a(x),\partial_y\theta_b(y)]=i\pi\,\delta_{ab}\delta(x-y),\notag\\
	\rho_{R,a}\!\pm\!\rho_{L,a}&=-\frac{1}{\pi}\partial_x\phi_a
	\ \ \text{and}\ \
	\frac{1}{\pi}\partial_x\theta_a.
\end{align}
Densities and interactions
\begin{align}
	\rho_a(x)&=n_{i,a}-\nu_a\nonumber\\
	&=-\frac{1}{\pi}\partial_x \phi_a
	+\frac{1}{2\pi\alpha}\Big[e^{i2k_{F,a}x}e^{-2i\phi_a}+h.c.\Big]\nonumber\\
\end{align}
where $\alpha$ is a cutoff and  $a=1,2,3$. The interaction becomes
\begin{align}
	H_{\text{int}}
	=U\!\int\!dx\,(\rho_1\rho_2+\rho_2\rho_3)
	+U'\!\int\!dx\,\rho_1\rho_3
	=H_{\text{fwd}}+H_{2k_F},
\end{align}
with
\begin{align}
	H_{\text{fwd}}=&\frac{U}{\pi^2}\!\int\!dx\,
	(\partial_x\phi_1\,\partial_x\phi_2+\partial_x\phi_2\,\partial_x\phi_3)\nonumber\\
	&+\frac{U'}{\pi^2}\!\int\!dx\,\partial_x\phi_1\,\partial_x\phi_3, \\
	H_{2k_F}=&\frac{2U}{(2\pi\alpha)^2}\!\int\!dx\,
	[\cos(2\phi_1-2\phi_2)+\cos(2\phi_2-2\phi_3)]\nonumber\\
	&+\frac{2U'}{(2\pi\alpha)^2}\!\int\!dx\,\cos(2\phi_1-2\phi_3).
\end{align}

Diagonalization and Luttinger parameters. Neglecting oscillatory terms $H_{2k_F}$:
\begin{align}
	H_L=\frac{1}{2\pi}\!\int\!dx\Big[
	(\partial_x \boldsymbol{\Phi})^T V_\phi (\partial_x \boldsymbol{\Phi})
	+(\partial_x \boldsymbol{\Theta})^T V_\theta (\partial_x \boldsymbol{\Theta})\Big],
\end{align}
with
\begin{align}
	&\boldsymbol{\phi}=(\phi_1,\phi_2,\phi_3)^T,\quad
	\boldsymbol{\theta}=(\theta_1,\theta_2,\theta_3)^T,\nonumber\\
	&V_{\phi}=v_F \mathbf{1}+\frac{1}{\pi}\mathbf{M},\quad
	V_{\theta}=v_F \mathbf{1},
\end{align}
\[
\mathbf{M}=
\begin{pmatrix}
	0 & U & U' \\
	U & 0 & U \\
	U' & U & 0
\end{pmatrix}.
\]
Eigenvalues/eigenvectors of M:
\begin{align}
	&m_1=-U',\qquad
	m_{2,3}=\frac{U' \pm \sqrt{U'^2+8U^2}}{2}, \nonumber\\
	&\mathbf e_1=\tfrac{1}{\sqrt2}(-1,0,1)^T,\ \
	\mathbf e_{2,3}=\frac{(1,\alpha_\pm,1)}{\sqrt{2+\alpha_\pm^2}},
\end{align}
\[
\alpha_{\pm}=\frac{-U'\pm\sqrt{U'^2+8U^2}}{2U},\qquad
K_{\sigma}=\Big(1+\frac{m_{\sigma}}{\pi v_F}\Big)^{-1/2}.
\]
For the total charge mode
\(\mathbf e_c=\frac{1}{\sqrt3}(1,1,1)^T\),
\begin{align}\label{kc-en}
	K_c=\sqrt{\frac{\mathbf e_c^T V_{\theta}\mathbf e_c}{\mathbf e_c^T V_{\phi}\mathbf e_c}}
	=\Big(1+\frac{2(U'+2U)}{3\pi v_F}\Big)^{-1/2}.
\end{align}
When \(U=U'\), this reduces to the \(SU(3)\) charge mode result
(valid for \(U\ll t\))\cite{assaraf1999metal}. For \(SU(2)\), any \(U>0\) opens a charge gap
(Mott insulator), while for \(SU(N>2)\) a finite \(U_c\) is required.

\subsection{Bosonizing \(\Delta(x)=c_1(x)c_3(x)\) and \(K_{\rm SC}\)}
Define \(\Delta(x)=c_1(x)c_3(x)\). The slow part comes from RL/LR.
For RL:
\begin{align}
	\Delta_{RL}=\psi_{R,1}\psi_{L,3}
	\propto e^{-i(\phi_1-\theta_1)}e^{+i(\phi_3+\theta_3)}
	=e^{\,i(\ell\cdot\Phi+\ell'\cdot\Theta)},
\end{align}
with \(\ell=(-1,0,1)\), \(\ell'=(1,0,1)\).
Project onto eigenmodes:
$\tilde\ell_\sigma=e_\sigma\!\cdot\!\ell,\quad\tilde\ell'_\sigma=e_\sigma\!\cdot\!\ell',$
giving $\tilde\ell_1=\sqrt2,\quad\tilde\ell'_1=0;\quad\tilde\ell_2=0,\tilde\ell'_2={2}/{\sqrt{2+\alpha_+^2}}; \quad
\tilde\ell_3=0,\ \tilde\ell'_3={2}/{\sqrt{2+\alpha_-^2}}.$ Scaling dimension for \(O=e^{i(\ell\cdot\Phi+\ell'\cdot\Theta)}\):

\begin{align}
\Delta[O]=\frac14\sum_{\sigma=1}^3
\left(\frac{\tilde\ell_\sigma'^2}{K_\sigma}
+K_\sigma\,\tilde\ell_\sigma^2\right).
\end{align}

Hence $\langle \Delta^\dagger(x)\, \Delta(0) \rangle \sim |x|^{-K_{\rm SC}}$.
\begin{align}
	K_{\rm SC}^{(0)} = 2\Delta[O] 
	= K_1 + \frac{2}{(2+\alpha_+^2)K_2} 
	+ \frac{2}{(2+\alpha_-^2)K_3}.
\end{align}

Using
$K_\sigma = (1 + {m_\sigma}/{\pi v_F} )^{-1/2}$
and $
2 + \alpha_\pm^2 = ({8U^2 + U'^2 \mp U' \sqrt{U'^2 + 8U^2}})/{2U^2},$
one finds, with \(v_F=2t\sin(\pi\nu)\),
\begin{align}
	&K_{\rm SC}^{(0)}(U,U';\nu)
	=\left(1-\frac{U'}{2\pi t\sin\pi\nu}\right)^{-1/2}\nonumber\\
	&+\frac{4U^2}{\,8U^2+U'^2-U'\sqrt{U'^2+8U^2}\,}
	\left(1+\frac{U'+\sqrt{U'^2+8U^2}}{4\pi t\sin\pi\nu}\right)^{1/2}\notag\\
	&+\frac{4U^2}{\,8U^2+U'^2+U'\sqrt{U'^2+8U^2}\,}
	\left(1+\frac{U'-\sqrt{U'^2+8U^2}}{4\pi t\sin\pi\nu}\right)^{1/2}.
\end{align}
Given $U=4, U^{\prime} = 0, \nu = 1/4$, we find $K_{SC} \approx 1.75 $ according to above equation.

	\bibliographystyle{apsrev4-2}
	\bibliography{SU3_1Dchain} 	
\end{document}